\documentclass[onecolumn]{emulateapj}

\usepackage{apjfonts}

\catcode`\@=11
\newcommand{\gapprox}{\mathrel{\mathpalette\@versim>}}
\newcommand{\lapprox}{\mathrel{\mathpalette\@versim<}}
\newcommand{\propapprox}{\mathrel{\mathpalette\@versim\propto}}
\newcommand{\@versim}[2]
  {\lower3.1truept\vbox{\baselineskip0pt\lineskip0.5truept
\ialign{$\m@th#1\hfil##\hfil$\crcr#2\crcr\sim\crcr}}}
\catcode`\@=12

\begin{document}


\title{Spectral Breaks in Pulsar-Wind Nebulae\\ 
and Extragalactic Jets}


\author{Stephen P. Reynolds}
\affil{Physics Department, North Carolina State University,
    Raleigh, NC 27695}


\begin{abstract}
Flows of synchrotron-emitting material can be found in several
astrophysical contexts, including extragalactic jets and pulsar-wind
nebulae (PWNe). For X-ray synchrotron emission, flow times are often
longer than electron radiative lifetimes, so the effective source size
at a given X-ray energy is the distance electrons radiating at that
energy can convect before they burn off.  Since synchrotron losses
vary strongly with electron energy, the source size drops with
increasing X-ray energy, resulting in a steepening of the synchrotron
spectrum.  For homogeneous sources, this burnoff produces the
well-known result of a steepening by 0.5 in the source's integrated
spectral index.  However, for inhomogeneous sources, different amounts
of steepening are possible.  I exhibit a simple phenomenological
picture of an outflow, with transverse flow-tube radius,
magnetic-field strength, matter density, and flow velocity all varying
as different powers of distance from the injection point.  For such a
picture, I calculate the value of the spectral index above the break
as a function of the power-law indices, and show the possible range of
steepenings.  I show that these simple calculations are confirmed by
full integrations of source luminosity, which also include the
spectral ``bump'' below the break from the accumulation of electrons
formerly at higher energies.  In many cases, extragalactic jets show
X-ray synchrotron emission steeper by more than 0.5 than the radio
emission; the same phenomenon is exhibited by many pulsar-wind
nebulae.  It is possible that source inhomogeneities are responsible
in at least some cases, so that the amount of spectral steepening
becomes a diagnostic for source dynamical or geometrical properties.

\end{abstract}


\section{Introduction}

\subsection{Spectral breaks in jets and pulsar-wind nebulae}

Broad-band spectra from synchrotron radiation characterize a wide
variety of astrophysical sources, including active galactic nuclei
(AGN), shell supernova remnants (SNRs), and pulsar-wind nebulae
(PWNe).  When observed over a sufficiently broad frequency range, such
spectra almost invariably show steepening to higher energies.  Such
steepening can be attributed either to intrinsic spectral structure or
to the effects of radiative losses.  However, the power-law shape of
radio spectra ($S_\nu \propto \nu^{-\alpha}$), with $\alpha \sim 0.5 -
0.8$ for many sources, suggests an origin of the requisite
relativistic electrons in diffusive shock acceleration (DSA), which
produces a power-law (or near-power-law, for efficient nonlinear DSA)
spectrum of particles $N(E) \propto E^{-s}$ with $s = 2\alpha + 1$
depending on the shock compression ratio.  Shocks are clearly present
in these astrophysical objects: outer blast waves in SNRs (and perhaps
reverse shocks into ejecta as well, for younger objects);
relativistic-wind termination shocks in PWNe; and shocks in jets and
hotspots in active galaxies.  The absence of an obvious mechanism to
generate a broken power-law distribution, and the necessity of the
operation of radiative losses at some level, has led to the common
acceptance of synchrotron losses as the mechanism to bring about
spectral steepening.

The standard calculation of the effect of synchrotron losses (reviewed
in the next section) for a homogeneous source predicts a steepening of
the initial electron spectrum to a second power-law one power steeper
($s_2 = s + 1$) than the injection spectrum, implying a radiation
spectrum one-half power steeper ($\alpha_2 = \alpha + 0.5$).  This is
in fact rarely observed.  In shell supernovae, the maximum energies to
which electrons can be accelerated are limited by losses or by finite
shock age (or size), but produce an exponential cutoff in $N(E)$,
further broadened by inhomogeneities, as observed in a few cases in
which synchrotron X-ray emission can be identified (see Reynolds 2008
for a review), so a sharp spectral break to a steeper power-law is
neither expected nor observed.  In PWNe, radio spectra are almost all
flatter than $\alpha = 0.5$, a phenomenon not well explained at
present, but the steepenings $\Delta \equiv \alpha_{\rm high} -
\alpha_{\rm low}$ are almost always greater than 0.5 (ranging from 0.7
to 1, in seven of the eight cases tabulated in Chevalier 2005, using
updated values for B0540-693 from Williams et al.~2008).  Knots in
extragalactic jets, when observable in X-rays, show similar too-large
steepenings (e.g., M87: Perlman \& Wilson 2005; Cygnus A: Stawarz et
al.~2007).  One standard interpretation of knot and hot-spot spectra
invokes shock acceleration and subsequent spectral steepening in a
uniform post-shock region \citep{heavens87}, but it cannot explain
these larger values of $\Delta$.  Coleman \& Bicknell (1988) present
numerical hydrodynamic simulations and find larger values of $\Delta$,
which they apply to observations, but without analytic results
allowing the wider application of the results.  While there is
recognition of the possibility of values of $\Delta$ different from
0.5 (e.g., Kennel \& Coroniti 1984a, who find $\Delta = (4 +
\alpha)/9$ for the Crab Nebula, or Petre et al.~2007, in a qualitative
discussion of the broadband spectrum of the PWN B0540-693), there has
as yet been no simple characterization of conditions under which
values of $\Delta \ne 0.5$ can naturally arise.  That characterization
is the goal of this paper.

\subsection{Synchrotron losses}

The first widely known calculation of the behavior of a distribution
of electrons subject to synchrotron losses is that of Kardashev
(1962).  While most of these results are well known, it is important
to recall the particular conditions under which each is applicable, so
I shall beg the reader's indulgence for a brief review, which can also
serve to fix notation.  Kardashev solved the kinetic equation for a
distribution $N(E,t)$ of electrons subject to gains by first and
second-order Fermi acceleration, and losses due to radiation or
adiabatic expansion, with and without the assumption of new-particle
injection and stationarity.  He writes the energy-loss rate from a
single electron as 
\begin{equation}
\dot E = -b B_\perp^2 E^2, 
\label{edot}
\end{equation}
where $B_\perp \equiv
B\sin\theta$, $b \equiv (2/3)(e^4/m_e^4 c^7)= 2.37\times 10^{-3}$ cgs
\citep[e.g.,][]{pacholczyk70}, and $\theta$ is the electron pitch
angle between its velocity vector and the magnetic field.  Kardashev
pointed out that in a uniform magnetic field in the absence of
scattering, electrons preserve their value of $\theta,$ since they
radiate a beam pattern that is symmetric with respect to their
velocity vector.  An electron injected into $B$ at $t = 0$ with energy
$E_0$ has an energy $E$ after time $t$ given by the well-known result
\begin{equation}
E(t) = {E_0 \over 1 + E_0 b B_\perp^2 t}.
\label{eoft}
\end{equation}
An initially infinitely energetic electron is reduced after time $t$
to energy $E_{\rm max}(t, \theta) = 1/b B_\perp^2 t$.  A power-law
energy distribution of electrons $N(E_0) = K E_0^{-s}$, with a single
value of pitch angle, will be cut off at $E_{\rm max}(t, \theta)$.  
The electron distribution $N(E)$ will evolve according to
$N(E)dE = N(E_0)dE_0$, so that 
\begin{equation}
N(E) = K \left[E_0(E)\right]^{-s} {dE_0 \over dE},
\label{evoldist}
\end{equation}
with $E(E_0)$ given by Equation~\ref{eoft}.  If $s < 2$, the electrons
initially above $E_{\rm max}(t)$ are sufficiently numerous to pile up
in a ``bump'' just below $E_{\rm max}(t)$, while if $s > 2$, the
``bump'' disappears.  (Numerical solutions to Equation~\ref{evoldist}
are shown below for inhomogeneous models, illustrating the ``bump''
effect.)

An initially isotropic distribution of electrons suffers unequal
radiation losses, with electrons with large pitch angles being more
rapidly depleted.  For an initial isotropic power-law distribution,
after a time $t$ one finds no electrons with pitch angles greater than
$\theta_{\rm max}$ given by $\sin^2 \theta_{\rm max} = 1/(b E B^2 t)$.
Since for synchrotron radiation, an individual electron's radiation
pattern has an angular width of order $1/\gamma$ where $\gamma$ is the
individual Lorentz factor, and $\gamma \gapprox 10^3$ for radio
emission and higher frequencies, we can approximate electrons as
radiating exactly in their directions of motion.  Then $\theta$ is
also the angle between the line of sight and the local magnetic field.
An initially isotropic distribution of electrons injected at $t = 0$
into a source with a uniform magnetic field making an angle $\chi$
with the line of sight, observed through its synchrotron emission,
would then disappear abruptly at time $t(\chi)$.  More realistically,
one might expect that the source has a tangled magnetic field, so that
all values of $\chi$ are achieved in some part of the source.  One
should then (for an unresolved source) perform an angular integration
over the electron distribution. The result is an electron distribution
that steepens by one power of $E$, i.e., $N(E) \propto E^{-s-1}$,
above the characteristic energy $E_b = 1/(aB^2 t)$.  (The synchrotron
emission from this distribution steepens above $\nu(E_b)$ by more than
the value of 0.5 in spectral index $\alpha$ ($S_\nu \propto
\nu^{-\alpha}$) because, in this time-dependent case without
continuous injection, a correlation exists between $E$ and $\theta$
such that more efficiently radiating electrons are depleted most
rapidly.)

This situation is still relatively unrealistic, as it ignores any
processes by which electrons could change their pitch angles.  If
electrons scatter in pitch angle on timescales much shorter than the
synchrotron-loss timescale, one should simply average the energy-loss
rate over angles: 
\begin{equation}
\dot E = 1.57 \times 10^{-3} B^2 E^2 \equiv a B^2 E^2, 
\label{edota}
\end{equation}
where $a \equiv b<\sin^2 \theta> = (2/3)b$.  Then the electron
distribution will remain isotropic, and will simply cut off at $E_{\rm
max}(t)$; a source synchrotron spectrum would then cut off
exponentially above $\nu(E_{\rm max} (t)$.

However, the result we all remember from graduate school is neither of
these.  A source that turns on at $t = 0$ with continuous, uniformly
spatially distributed injection of a power-law distribution of
electrons $q(E) \equiv J_0 E^{-s}$ electrons erg$^{-1}$ s$^{-1}$
cm$^{-3}$, develops a break at energy $E_b = 1/(aB^2t)$, where the
electron spectrum steepens by one power in $s$.  This corresponds to a
steepening of the synchrotron spectrum by one-half power in $\alpha$
at $\nu_b = c_1 B^{-3} t^{-2}$ with $c_1 = 1.12 \times 10^{24}$ cgs.
This relation is frequently used to deduce a magnetic-field strength
in a synchrotron source of known age.

The result that synchrotron losses (or inverse-Compton losses, which
have the same dependence on electron energy) result in the steepening
of the particle spectrum by one power and the steepening of the
emitted synchrotron spectrum by a half-power, is a widely held idea.
It is this application that will be generalized below.  It is
important to remember that the standard derivation assumes a
distributed injection of electrons in a homogeneous source.  For other
conditions, it is not correct, as we shall see.

\section{Basic Calculation}

Energy-loss breaks from a synchrotron source in which relativistic
electrons are advected systematically away from an injection region
(such as a wind-termination shock) can be described as being due to
shrinking of the effective source size with frequency.  At a high
enough observing frequency, electrons drop below the energy required
to emit at that frequency before they reach the edge of the object,
hence limiting its effective size at that frequency.  Thus all results
depend on the critical energy an electron may have after suffering
synchrotron losses and convecting at speed $v$.  Equation~\ref{edota}
gives the synchrotron loss rate from a single electron.  In a
constant-density flow (i.e., neglecting adiabatic losses), but
allowing the possibility of spatially varying $B$, we generalize
the homogeneous results slightly to obtain
\begin{equation}
E_c = \left[\int a \,B^2\,{dr \over v}\right]^{-1}
\label{ec0}
\end{equation}
for the energy an initially infinitely energetic electron would have
after moving at $v$ through a field $B$ for a distance $r$.  So any
injected electron distribution must cut off at this energy.  For a
one-dimensional flow of electrons streaming at constant speed $v$ in a
constant magnetic field $B$, the effective length $r(E)$ of the source
is found from
\begin{equation}
E(r) = \left( {v \over a B^2} \right) r^{-1}.
\end{equation}
An electron of this energy radiates chiefly at frequency
$\nu = c_{\rm m} E^2 B = c_{\rm m}(v^2/a^2 B^3) r^{-2}$ where
$c_{\rm m} = 1.82 \times 10^{18}$ cgs (e.g., Pacholczyk 1970).
Then at any frequency $\nu$
the source has an effective length
\begin{equation}
r(\nu) = c_{\rm m}^{1/2} \left( v \over a B^{3/2} \right) \nu^{-1/2}.
\end{equation}
For synchrotron emission with a spectral index $\alpha$, the 
observed flux density would then vary as
\begin{equation}
S_\nu \propto \nu^{-\alpha} r(\nu) \propto \nu^{-\alpha-1/2}
\end{equation}
-- the famous steepening by one-half power in the spectral index.
(Note that at no position $\bf {r}$ in the source is there a break in
the electron energy distribution $N({\bf r}, E)$ to a new power-law,
although such a distribution does result after integrating over
the entire source volume to obtain $N(E)$.)

This kind of argument can be generalized to inhomogeneous sources, for
which the steepening may be larger or smaller than 0.5.  For instance,
if the synchrotron emissivity increases with distance from the center,
then as the effective source size shrinks, more emission will be lost
than if the source were homogeneous, and the steepening can be greater
than 0.5.

\begin{figure}
\epsscale{0.80}
\plotone{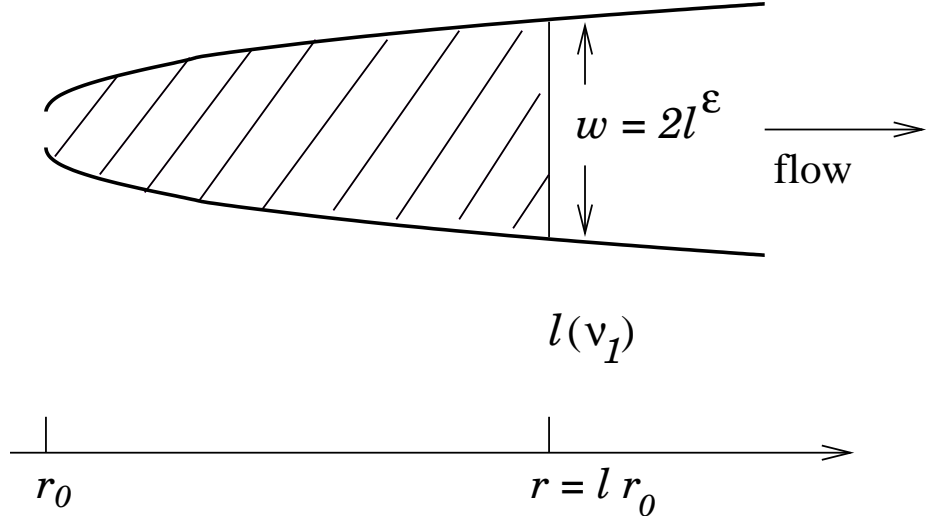}
\caption{Schematic of flow geometry.  The flow occurs in a tube
whose cross-sectional width $w$ grows as a power $\epsilon$ of (normalized)
distance $l$ from the injection radius $r_0$ ($l \equiv r/r_0$).
\label{jetfig}}
\end{figure}

Here we consider a simple model in which electrons are injected at
some initial radius $r_0$ in a ``jet'' whose full width $w$ rises with
dimensionless distance $l \equiv r/r_0$ as $l^\epsilon$: $w =
w_0 l^\epsilon.$ A conical jet (or piece of spherical outflow) then has
$\epsilon = 1$; a confined jet has $\epsilon < 1.$ Then the
cross-sectional area increases as $A_\perp \propto l^{2\epsilon}$ (see
Figure~\ref{jetfig}).  We shall assume ad-hoc power-law dependences of
quantities on dimensionless length $l$, ignoring any transverse
variations -- so a one-dimensional problem.  Let the $l$-dependence
of basic quantities be given by
\begin{equation}
B = B_0 l^{m_b} \qquad v = v_0 l^{m_v} \qquad \rho = \rho_0 l^{m_\rho}.
\end{equation}
While this is completely general, physical constraints may couple the
$m$'s.  For instance, in the absence of some mechanism such as mass
loading or entrainment \citep{lyutikov03} to alter the effective
density $\rho$, conservation of mass gives
\begin{equation}
{\rm Mass \ conservation}\ \qquad  \rho v A_\perp = {\rm const} 
    \Rightarrow m_\rho + m_v = -2\epsilon.
\label{masscons}
\end{equation}
In the absence of turbulent amplification or reconnection, magnetic flux
will be conserved, giving different relations for components of
$B$ parallel and perpendicular to the jet axis:
\begin{eqnarray}
{\rm Longitudinal \ (radial) \ field}\ \qquad  B_\parallel: BA_\perp = {\rm const} 
    \Rightarrow & m_b = -2\epsilon\\
{\rm Transverse \ (toroidal) \ field} \ \qquad B_\perp: Bwv = {\rm const} 
    \Rightarrow & m_b = -m_v - \epsilon = m_\rho + \epsilon
\label{fluxcons}
\end{eqnarray}
where the last form for transverse $B$ involves invoking mass conservation.
Of course, unless the magnetic field is exactly radial, any toroidal component
will eventually dominate, barring extremely peculiar and probably unphysical
behaviors (e.g., $m_\rho < -3$).

In this formalism, Equation~\ref{ec0} implies $E_c \propto l^{-(1 +
2m_b - m_v)}$ in the absence of other energy-loss mechanisms such as
adiabatic expansion losses (non-constant density). In the presence of
adiabatic losses, it can be shown
\citep[e.g.,][]{kennel84a,reynolds98} that the critical energy is
given by
\begin{equation}
E_c = \rho^{1/3}\left[ \int_1^l {a B^2 \over v} \rho^{1/3} r_0 dl\right]^{-1} = 
l^{m_\rho/3}
\left[ \int_1^l a v_0^{-1} r_0 B_0^2 l^{2m_b -m_v + m_\rho/3} dl \right]^{-1}
\label{ec1}
\end{equation}
\begin{equation}
={v_0 \over ar_0 B_0^2} (1 + 2m_b - m_v + m_\rho/3)
l^{-(1 + 2m_b - m_v)}
\equiv A_E l^{m_E}
\end{equation}
where we have assumed the source is long enough that $l(E) \gg 1$.
Note that the effects of adiabatic losses have canceled out (except
for a small change in the integration constant), leaving $E_c$ with
the same $l$-dependence as in the constant-density case.  We have also
made the assumption that the integral in Equation~\ref{ec1} is
dominated by the upper limit at $l$, demanding that $m_c \equiv 2m_b
-m_v + m_\rho/3 > -1$.  Otherwise, $E_c \propto l^{m_\rho/3}$ and
radiative losses play no role in the spectral behavior, so that the
calculation is not self-consistent.  Even if this condition is met, we
still wish to exclude situations in which the gradients conspire to
arrange adiabatic {\sl gains} of electrons as they convect, i.e., $m_E
> 0$.  This situation appears both unphysical and unlikely.  Note that
$m_c = m_\rho/3 - m_E - 1$.  These two conditions rule out some volume
in the parameter space of $(\epsilon, m_i)$ and must be checked for
any particular choices of those parameters.  The constraints are
related; $m_c > -1$ requires $m_E < m_\rho/3$, so the we ultimately
require $m_E < {\rm min}(0, m_\rho/3)$.

Electrons with energy $E_c$ (at position $l$ where the magnetic
field strength is $B(l)$) radiate chiefly at a frequency
\begin{equation}
\nu(E_c) = c_{\rm m} \left( v_0^2 \over a^2 r_0^2 B_0^3\right)
  \left(1 + 2m_b - m_v + m_\rho/3\right)^2
l^{m_b-2(1 + 2m_b - m_v)}
\equiv A_\nu l^{m_\nu}
\label{nub}
\end{equation}
defining $A_\nu$ and $m_\nu$.  Note $m_{\nu} = m_b + 2m_E = -2 - 3m_b
+ 2m_v$.  Then
\begin{equation}
l(\nu) = A_\nu^{-1/m_\nu} \nu^{1/m_\nu} \equiv A_l \nu^{1/m_\nu}.
\label{lnu}
\end{equation}
We are focusing on conditions such that the source size shrinks with
increasing frequency, i.e., $m_\nu < 0$.  It can easily be shown that
the condition on $m_c$ above is equivalent to $m_\nu < (m_\rho + m_v)
-1$.  If mass conservation is assumed, $m_\rho + m_v = -2\epsilon$ and
$m_\nu < 0$ always.  Otherwise, this condition must be checked, but
for reasonable values of the parameters (such as those in all examples
described below) it is always fulfilled.

Now to discuss the synchrotron flux, we assume a power-law electron
distribution $N(E) = KE^{-s}$ between energies $E_l$ and $E_h \gg E_l$
(we take $s > 1$).  As the flow expands, conservation of electron
number gives
\begin{equation}
n_e = {K \over s-1} E_l^{1-s} \propto \rho \Rightarrow
  K \propto \rho E_l^{s-1} \propto \rho^{1 + (s-1)/3} = \rho^{(s+2)/3}
\equiv \rho^{(2\alpha + 3)/3}
\end{equation}
since adiabatic losses give individual-particle energies varying as $E
\propto \rho^{1/3}$, at least near $E_l$, an energy we assume to be too low
for radiative losses to be important.  We have taken $s$ enough larger
than 1 that $E_l^{1-s} \gg E_h^{1-s}$.
Then we can write the synchrotron emissivity
(following Pacholczyk 1970) as
\begin{equation}
j_\nu = c_j K B^{1+ \alpha} \nu^{-\alpha}
= c_j K_0 B_0^{1+\alpha} l^{(2\alpha + 3)m_\rho/3 + (1 + \alpha)m_b} \nu^{-\alpha}
\equiv A_j l^{m_j} \nu^{-\alpha}.
\end{equation}
Here $c_j(s) \equiv c_5(s) (2c_1)^\alpha$ in the notation of
Pacholczyk; for $s = 1.5$, $c_j = 1.34 \times 10^{-18}$ cgs, and we
have defined another important index, $m_j \equiv (2\alpha +
3)m_\rho/3 + (1 + \alpha)m_b$.  Assume for the time being that we view
the jet directly perpendicular to the axis.  Then the line-of-sight
path length through the jet at any position $l$ is just $w(l) = w_0
l^{\epsilon}$ (through the center; averaged over lines of sight
intersecting a circular cross-section, we obtain the mean
line-of-sight path of $(\pi/4)w$.  We recall that we are assuming all
jet quantities to be constant in cross-section, that is, along these
lines of sight.  So if the source is at distance $D$, the integrated
flux density $S_\nu$ is given by
\begin{equation}
S_\nu = \int I_\nu d\Omega 
= {1 \over D^2}\int_1^{l_\nu} dA \int_0^w {\pi \over 4} j_\nu ds 
= {\pi A_j \over 4 D^2}\int_1^{l_\nu} w \ r_0 dl (w \ l^{m_j})\nu^{-\alpha}
\label{flux1}
\end{equation}
\begin{equation}
={\pi A_j \over 4 D^2}\int_1^{l_\nu} r_0 w_0^2 \ l^{2\epsilon + m_j} \nu^{-\alpha}\ dl
={\pi A_j r_0 w_0^2 \over 4 D^2(1 + 2\epsilon + m_j)} 
    l_\nu^{1 + 2\epsilon + m_j}\nu^{-\alpha}
\label{flux2}
\end{equation}
%
%
\begin{equation}
={\pi c_j K_0 B_0^{1+\alpha}r_0 w_0^2 \over 4 D^2 
  \left[1 + 2\epsilon + (2\alpha + 3)m_\rho/3 + (1 + \alpha)m_b\right]}
A_l^{(1 + 2\epsilon + (2\alpha + 3)m_\rho/3 + (1 + \alpha)m_b)}
\nu^{- \alpha - \Delta}
\label{flux3}
\end{equation}
where the last equation defines $\Delta$, the amount of spectral
steepening:
\begin{equation}
\Delta = - {1 + 2\epsilon + m_j \over {m_b + 2m_E}} = 
  {1 + 2\epsilon + m_j \over |m_\nu|} = 
{1 + 2\epsilon + (2\alpha + 3)m_\rho/3 + (1 + \alpha)m_b
\over 2 + 3m_b -2m_v}. 
\label{Delta}
\end{equation}
We have assumed that $m_\nu < 0$, and that the flux integral
depends on the outer, not the inner, limit of integration, that is,
that
\begin{equation}
1 + 2\epsilon + m_j > 0.
\end{equation}
This latter condition can be restrictive.

While we have assumed a jet seen from the side, since the emission is
optically thin the observed flux density should be independent of
viewing angle.  This can be shown explicitly for the case of jets seen
exactly end-on, for which the flux integral Equation~\ref{flux1}
becomes
\begin{equation}
S_\nu = {1 \over D^2} \int_0^{w_{\rm max}} {\pi \over 2} w dw
  \int_{l_i(w)}^{l(\nu)} c_j K_0 B_0^{1 + \alpha}r_0 \nu^{-\alpha} l^{m_j} dl.
\label{fluxend}
\end{equation}
Here the line-of-sight is parallel to the jet axis, beginning
at a value of $l \equiv l_i$ dependent on $w$ (if $w < w_0, l_i = 1$),
The upper limit $w_{\rm max}$ is just the value of $w$ at which
$l_i = l_\nu$, i.e., $w_{\rm max} = w(l(\nu)) \equiv [l(\nu)]^\epsilon$.
Now we need a slightly more restrictive condition for the emission
to be dominated by the outer limit of integration $l(\nu)$:
$1 + m_j > 0$.  If this is the case, the two integrals in 
Equation~\ref{fluxend} decouple:
\begin{equation}
S_\nu = {\pi \over 2 D^2} c_j K_0 B_0^{1 + \alpha} r_0 \nu^{-\alpha}
  {w_{\rm max}^2 \over 2} {1 \over 1 + m_j} \left[ l(\nu) \right]^{1 + m_j}.
\end{equation}
But $w_{\rm max} = [l(\nu)]^\epsilon$, so
\begin{equation}
S_\nu \propto \left[ l(\nu)\right]^{1 + 2\epsilon + m_j} \nu^{-\alpha}
\propto \nu^{((1 + 2\epsilon + m_j)/m_\nu) - \alpha}
\end{equation}
just as in Equation~\ref{flux2}. 

These power-law spectra can hold only over a frequency range related
to the size range of the source by $l(\nu)$.  For instance, for
conical, constant-velocity, mass-conserving flow with tangential
magnetic field, we have $\epsilon = 1$, $m_\rho = -2$, and $m_b = -1$,
giving $l(\nu) \propto \nu^{-1}.$ Thus a source showing the
corresponding value of $\Delta$ (in this case, $\Delta = 7\alpha/3$)
between frequencies $\nu_1$ and $\nu_2$ must shrink over a range of
sizes given by $r_1/r_2 = \nu_2/\nu_1$.  In general, sources whose
spectra are set by effective variations of size with frequency are
predicted to have sizes varying as
\begin{equation}
l(\nu) \propto \nu^{1/m_\nu} \equiv \nu^{1/(-2 - 3m_b + 2m_v)}.
\label{size}
\end{equation}
Equation~\ref{Delta} relates the size exponent $1/|m_\nu|$ to
$\Delta$: $1/|m_\nu| = \Delta/(1 + 2\epsilon + m_j)$.  (Thus a source
with observed $\Delta$ will have a smaller rate of shrinkage with
frequency for a larger value of $1 + 2\epsilon + m_j$.)  The source
subtends a solid angle on the sky of
\begin{equation}
\Delta \Omega = {1 \over D^2} \int_1^{l(\nu)} (w_0 l^\epsilon) (r_0 dl)
= {1 \over D^2} {r_0 w_0 \over 1 + \epsilon} \left[l(\nu)^{1 + \epsilon}\right].
\label{solidang}
\end{equation} 
If the source is only marginally resolved, one may consider an ``average''
angular size $\langle\theta\rangle$ defined by
\begin{equation}
\langle\theta\rangle \equiv (\Delta \Omega)^{1/2} 
= {\sqrt{r_0 w_0} \over D} {1 \over \sqrt{1 + \epsilon}} [l(\nu]^{(1 + \epsilon)/2}
\end{equation}
so that $\langle\theta\rangle \propto \nu^{(1 + \epsilon)/2m_\nu}$.
For spherical or conical flows, i.e., $\epsilon = 1$,
$\langle\theta\rangle \propto \nu^{1/m_\nu}$ as before, but for
confined flows ($\epsilon < 1$), the average angular size decreases
more slowly with frequency.  If a confined jet is seen end-on, the
size variation with frequency is reduced even further, as the apparent
diameter is now proportional to $w_{\rm max} \propto
[l(\nu)]^\epsilon$ instead of $l(\nu)$:
\begin{equation}
\theta \propto \nu^{\epsilon/m_\nu} = \nu^{\epsilon/(m_b + 2m_E)}
   =\nu^{\epsilon/(-2 - 3m_b + 2m_v)}.
\end{equation}
This may result in a significantly reduced size effect, since the
primary change in emitting volume is a shrinking along the line of sight.

\section{Special Cases}

We can examine a few special cases.  First, in the case of a plane
constant-velocity flow, we have $\epsilon = 0, m_\rho = 0$, and $m_b =
0$, and we recover $\Delta = 1/2$.  Next, consider tangential magnetic
field, and conical outflow ($\epsilon = 1$) with constant density and
assuming mass and flux conservation. This corresponds to the inner
parts of a Kennel \& Coroniti (KC) MHD spherical flow.  Then $m_\rho =0$,
$m_v = -2$, $m_b = m_\rho + \epsilon = 1$, and we have
\begin{equation}
\Delta = {4 + \alpha \over 9}
\end{equation}
which was derived in Kennel \& Coroniti (1984b, eq.~4.11b).  This
already indicates that values of $\Delta \ne 0.5$ can be obtained
from reasonable situations.  It also suggests that obtaining values
very different from $0.5$ may be difficult.  The size effect predicted
above is quite weak:  $m_\nu = -2 - 3m_b + 2m_v = -9$, so $l(\nu)
\propto \nu^{-1/9}$, as shown in Kennel \& Coroniti (1984b, eq.~4.10b).
This weak effect accounts for the very small deviation of $\Delta$
from the homogeneous value of 0.5.  

Now KC models can be divided into two regions: an inner one as above,
and an outer one, a constant-velocity flow with $\rho \propto r^{-2}$
(that is, $m_v = 0$ and $m_\rho = -2$) and tangential field ($m_b =
m_\rho + 1 = -1$).
Mass conservation is assumed ($m_v = -2\epsilon -
m_\rho$).  At lower frequencies, the burnoff radius moves in through the
outer region, giving
\begin{equation}
\Delta_{\rm outer} = {7 \alpha \over 3}.
\end{equation}
However, this is not realized actually, as neither 
condition $m_E < 0$ or $1 + 2\epsilon + m_j > 0$ is met.  First, $m_E
\equiv -(3 + 2m_b + m_\rho) = + 1$, indicating that $E_c$ rises
with $l$.  Further, the $m_j$ condition gives $3 -2(2\alpha +
3)/3 -1 - \alpha = -7\alpha/3$, so the flux is dominated by the inner
parts of the nebula.  At high enough frequencies or photon energies
that the burnoff radius is at the transition point and moves into the
$v \propto r^{-2}$ region, we obtain the above result
\begin{equation}
\Delta_{\rm inner} = {4 + \alpha \over 9}.
\end{equation}
Here, $1 + 2\epsilon + m_j = 4 + \alpha > 0$, so the consistency condition
is met -- the flux is dominated by regions near $l(\nu)$.
For KC's model of the Crab, $\alpha = 0.6$, so $\Delta_{\rm inner} = 0.51.$  

It is straightforward to show directly that spherically symmetric
sources obey the same scaling laws with $\epsilon = 1$ -- that is,
the assumptions of thin jets perpendicular to the line of sight 
still give the correct scaling for spheres.  Let the dimensionless
radius be $R \equiv r/r_0$, with injection at $R = 1$ and burnoff at
\begin{equation}
R(\nu) = A_R \nu^{1/m_\nu} \equiv A_R \nu^{1/(-2 -3m_b + 2m_v))}
\end{equation}
-- that is, the same expression as for $l(\nu)$. 
(Sphericity shouldn't change the expression, only, perhaps, the values
of the $m$'s.)  Similarly, we should have the same dependence of
$j_\nu(R)$ in the spherical case as we had for $j_\nu(l)$ in the jet
case:
\begin{equation}
j_\nu(R) = c_j K_0 B_0^{1+\alpha} \nu^{-\alpha} R^{m_j}
\end{equation}
with
\begin{equation}
m_j \equiv {2\alpha + 3 \over 3}m_\rho + (1 + \alpha) m_b.
\end{equation}
Then the total flux from this spherically symmetric, optically thin
source is just
\begin{equation}
S_\nu = {4\pi \over D^2} \int_1^{R(\nu)} 4\pi j_\nu(R) R^2 dR
= {4\pi \over D^2}\left(4\pi c_j K_0 B_0^{1+\alpha}\nu^{-\alpha}\right)
{[R(\nu)]^{m_j + 3} \over {m_j +3}}
\end{equation}
which gives
\begin{equation}
\Delta \equiv - {m_j + 3 \over m_\nu} = 
+ {(2\alpha + 3)m_\rho/3 + (1+\alpha)m_b + 3  
  \over 2 + 3m_b - 2m_v}.
\end{equation}
This is the same result as can be obtained by setting $\epsilon = 1$ in
the previous expression for $\Delta$, Equation~\ref{Delta}.

\section{Breaks greater than 0.5}

Obtaining values of $\Delta$ considerably greater than 0.5, as seems
to be required by observations of most PWNe, requires relaxing some of
the assumptions.  The relation from mass conservation of $m_\rho =
-2\epsilon - m_v$ might not hold if some form of mass-loading occurs,
for instance by evaporation of material from thermal filaments, or
entrainment of material from a confining medium.  Flux-freezing for
the magnetic field will not hold in the presence of either turbulent
amplification of magnetic field or of reconnection.  As an example,
consider (for algebraic simplicity) the case $\alpha = 0$, but
conical, constant-density flow (so $m_v = -2$), still conserving mass.
Now in this case, $m_j = m_b$, and $m_E = -(3 + 2m_b)$, so
\begin{equation}
\Delta = - {{3 + m_b} \over {m_b -2(3 + 2m_b)}} = {3 + m_b \over 6 + 3m_b}
\ \Rightarrow \ m_b = -{6 \Delta - 3 \over 3\Delta - 1}.
\label{mb01}
\end{equation}
Now we can obtain $\Delta = 2/3$ with the value $m_b = -1$ in this
case: the magnetic field drops as the first power of distance (faster
than if frozen-in and tangential, slower than if longitudinal).  This
does not seem like an unreasonable possibility.  Note that the 
conditions are met: $1 + 2\epsilon + m_j = 3 + m_b = 2 > 0$,
$m_E = -1 < 0$, and $m_c \equiv 2m_b - m_v + m_\rho/3 = 0 > -1$.
The source size would obey
\begin{equation}
l(\nu) \propto \nu^{1/(m_b + 2m_E)} = \nu^{1/(-1 - 2(1))} = \nu^{-1/3}.
\end{equation}
To obtain weaker size effects, one is driven to larger values of
$1 + 2\epsilon + m_j$, for the same observed $\Delta.$

For practical use, it is convenient to consider $m_b$ a dependent
variable, in terms of the other quantities.  First, solve the equation
for $\Delta$ for $m_b$, in terms of $\Delta$, $m_\rho$, $\epsilon$,
and $\alpha$, with no presumed relation between $m_b$ and $m_\rho$.
The result is
\begin{equation}
m_b = {\Delta(2m_v - 2) + (2\alpha + 3)m_\rho/3 + 1 + 2\epsilon
   \over 3\Delta - 1 - \alpha}.
\label{mb}
\end{equation}
Two of the conditions are satisfied if $m_E < {\rm min}(0, m_\rho/3)$, or
\begin{equation}
1 + 2m_b - m_v > {\rm max}(0, -m_\rho/3).
\end{equation}
Assuming mass conservation, this becomes
\begin{equation}
1 + 2\epsilon + 2m_b + m_\rho > {\rm max}(0, -m_\rho/3).
\label{condition}
\end{equation}
In addition, we still require $1 + 2\epsilon + m_j > 0.$

\begin{figure}
\epsscale{1.0}
\plottwo{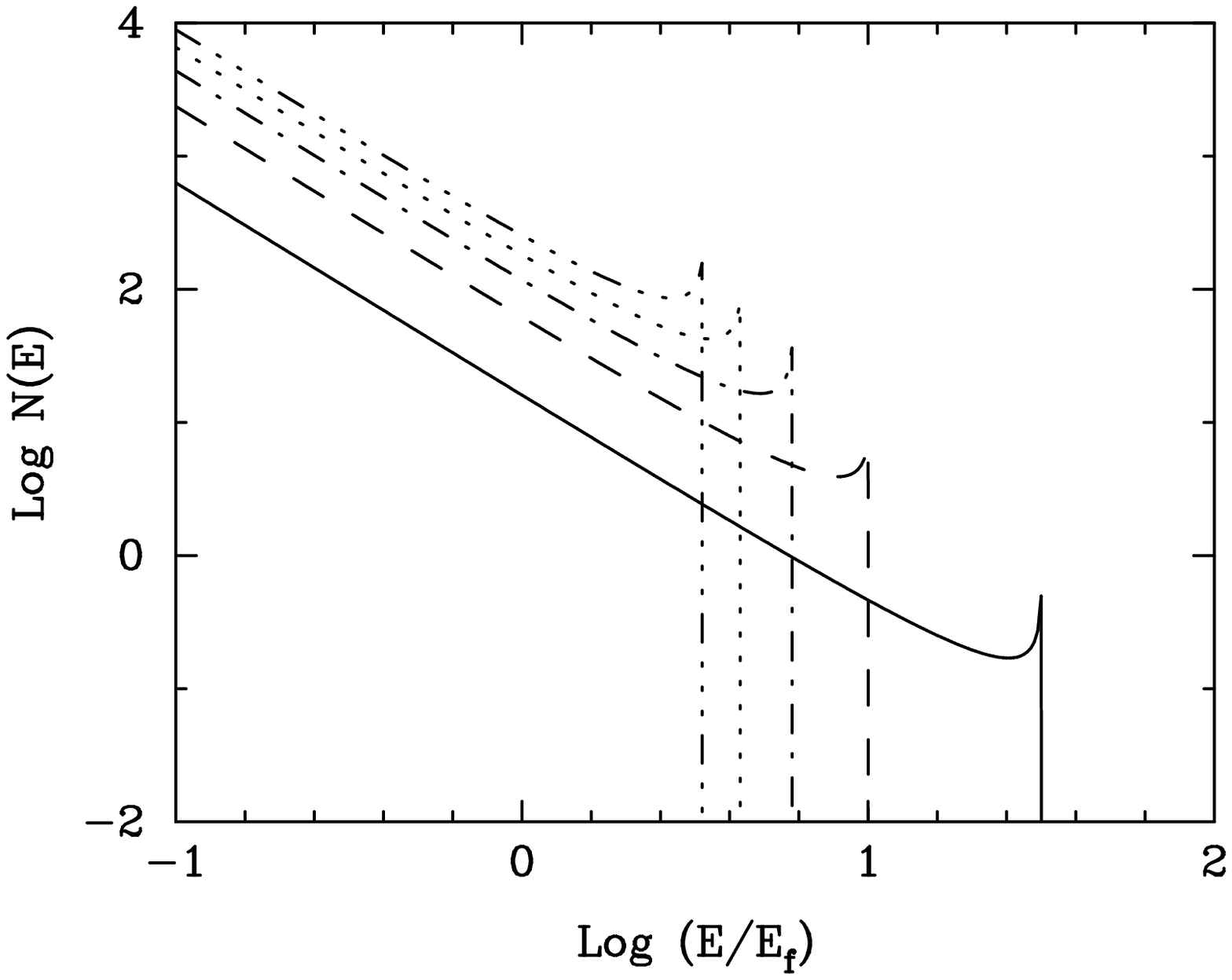}{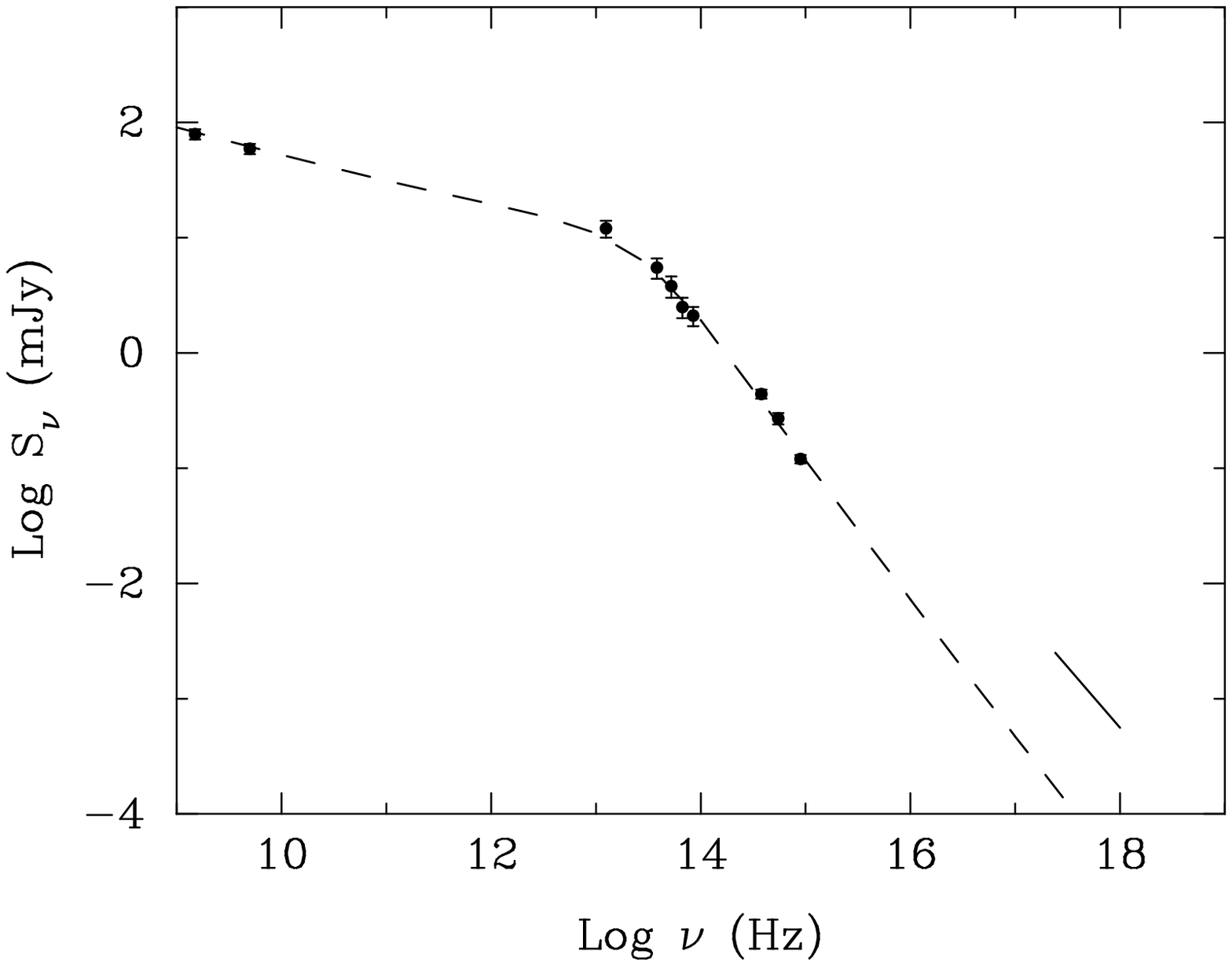}
\caption{Left: Electron distribution $N(E, l)$ at five positions in
the flow model for the PWN B0540-693: $l \equiv r/r_0 = 10, 30, 50,
70,$ and $90$.  Right: Model spectral-energy distribution and
observations for B0540-693, reproduced from Williams et al.~2008.
Radio: Manchester et al.~1993.  IR: Williams et al.~2008. Optical:
Serafimovich et al.~2004.  X-ray: Kaaret et al.~2001.
\label{b0540}}
\end{figure}

An application to a particular source (i.e., an object of known
$\alpha$ and $\Delta$, presumably) can then be made by inserting those
values.  Various possibilities for $\epsilon$ and $m_\rho$ can be
tried; for each value of $m_b$ obtained in this way, the condition on
$m_E$ must be checked by hand.  For example, consider B0540-693, with
a radio spectrum of $\alpha = 0.25$ (Manchester et al. 1993), and a
break with $\Delta \sim 1$ at around 20 $\mu$m \citep{williams08}.
A constant-density spherical (or conical) outflow cannot produce this
$\Delta$.  Inserting the values of $\alpha$ and $\Delta$ into
Equation~\ref{mb}, and assuming for simplicity $\epsilon = 1$,
we find
\begin{equation}
m_b = {1 + 2m_v + 1.2m_\rho \over 1.7}.
\end{equation}
Then at the price of abandoning mass conservation, we can choose
$m_\rho = 1$ and $m_v = -2$, giving $m_b = -1.06,$ or about $-1$.
The consistency conditions are all met:  $m_c = 1/3 > -1$, 
$m_E \equiv -(1 + 2m_b - m_v) = -1 < 0$,, and $1 + 2\epsilon + m_j
= 2.9 > 0$.   
The source effective radius decreases as
\begin{equation}
r \propto \nu^{1/(m_b + 2m_E)} = \nu^{-0.33}
\end{equation}
which may be a serious problem, since we need the slope of $-\alpha -
\Delta \cong -1.2$ to hold from about 20 $\mu$m to somewhere in the blue
or near UV -- say 0.2 $\mu$m, requiring that the source shrink between
these two wavelengths by a factor of 4.6 -- perhaps unlikely.  (Though
what is shrinking is really the region containing the dominant flux;
there could be a faint halo contributing a small amount of flux in
which a brighter, shrinking core is embedded).  A numerical
calculation of this model is shown in Figure~\ref{b0540}, along with
observations.  The physics which could cause these values of $m_\rho$
and $m_b$ is, of course, completely unknown.

\begin{figure}
\epsscale{1.0}
\plottwo{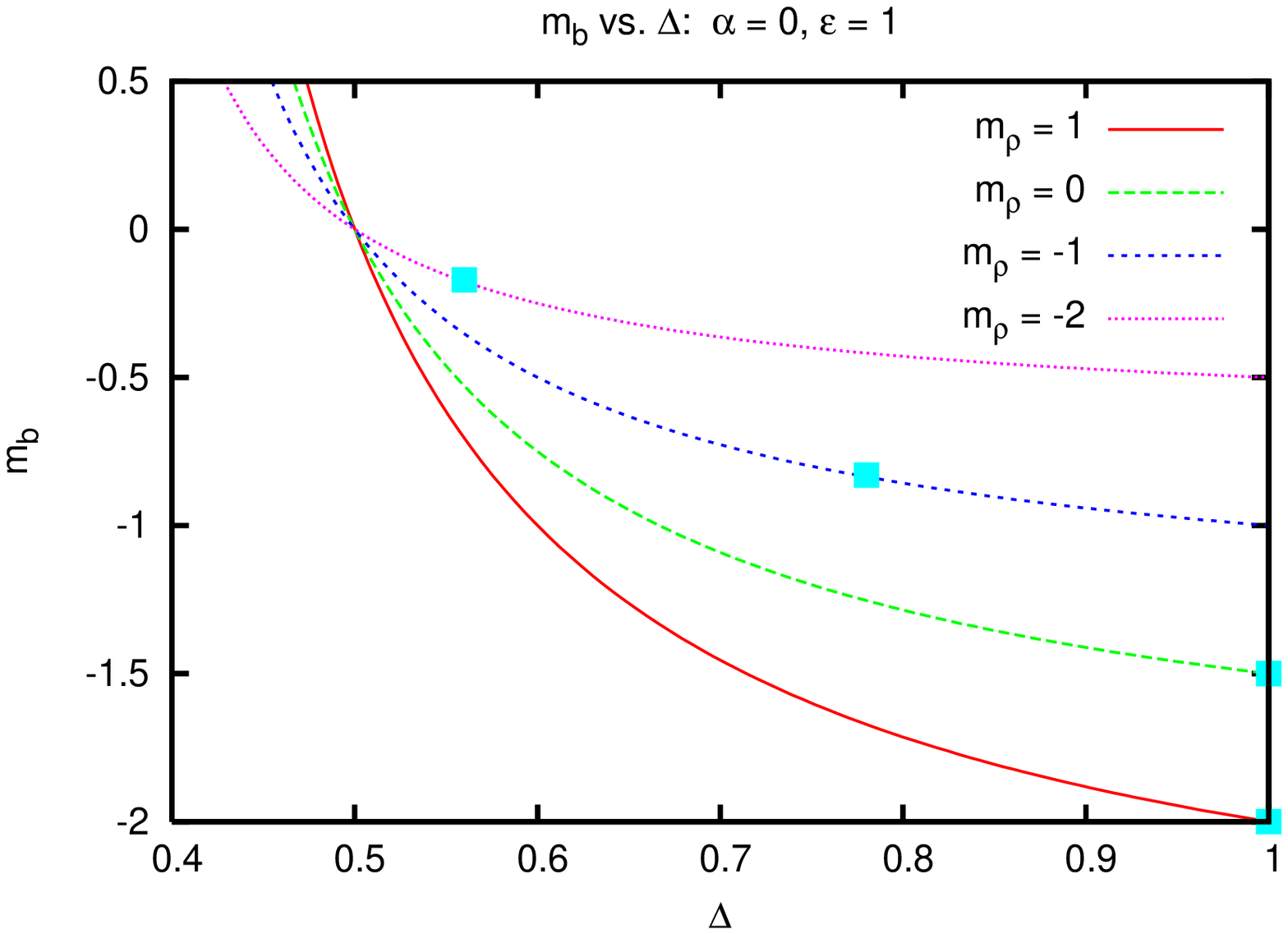}{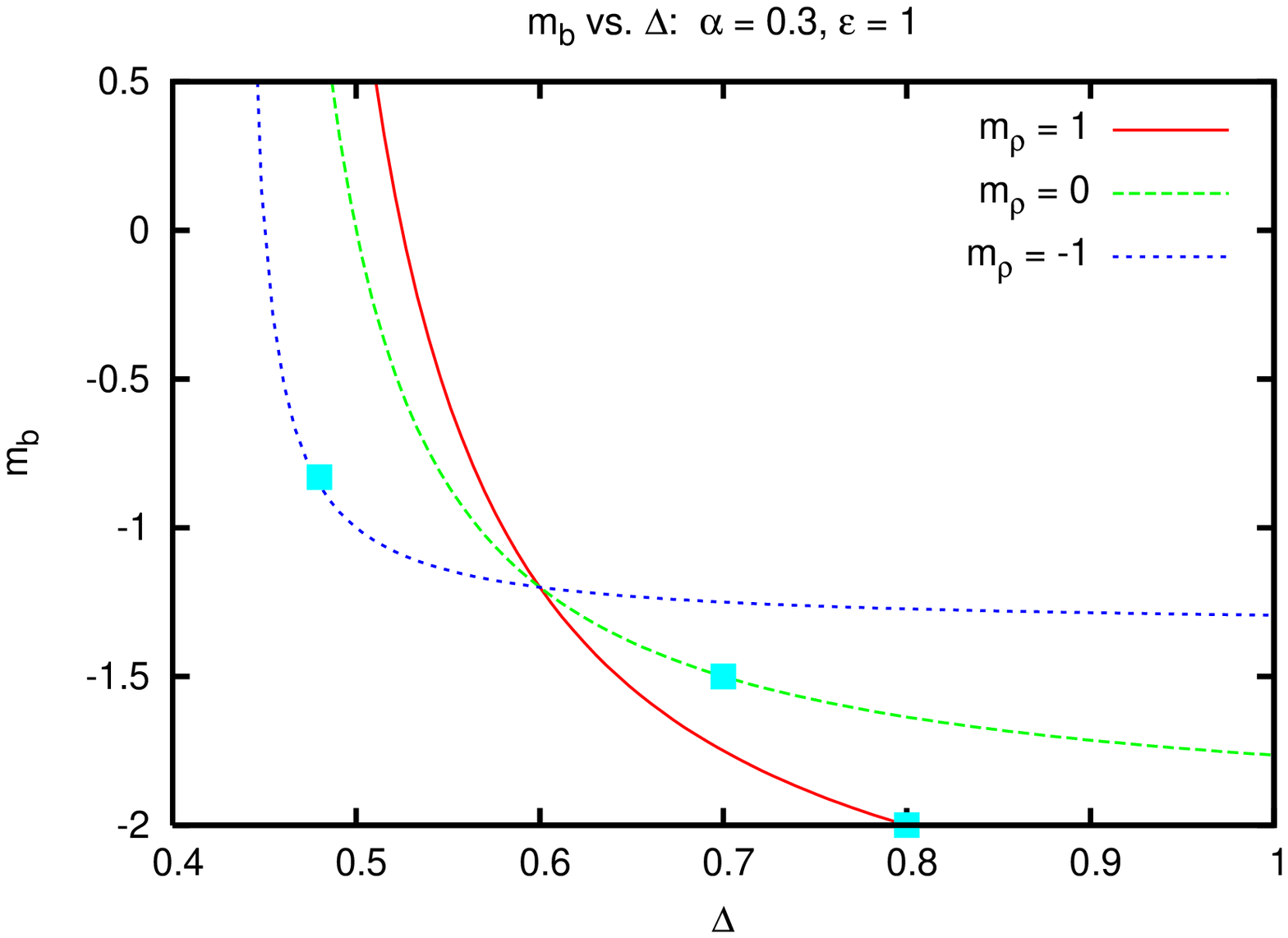}
\caption{Left:  magnetic-field index vs.~$\Delta$ for $\alpha = 0$, 
$\epsilon = 1,$ for several values of $m_\rho$.  Mass conservation
is assumed. Right:  Same, for $\alpha = 0.3$.  Only values of
$\Delta$ less than the square symbols on each curve satisfy the consistency
requirement.
\label{mb1}}
\end{figure}

\begin{figure}
\epsscale{1.0}
\plottwo{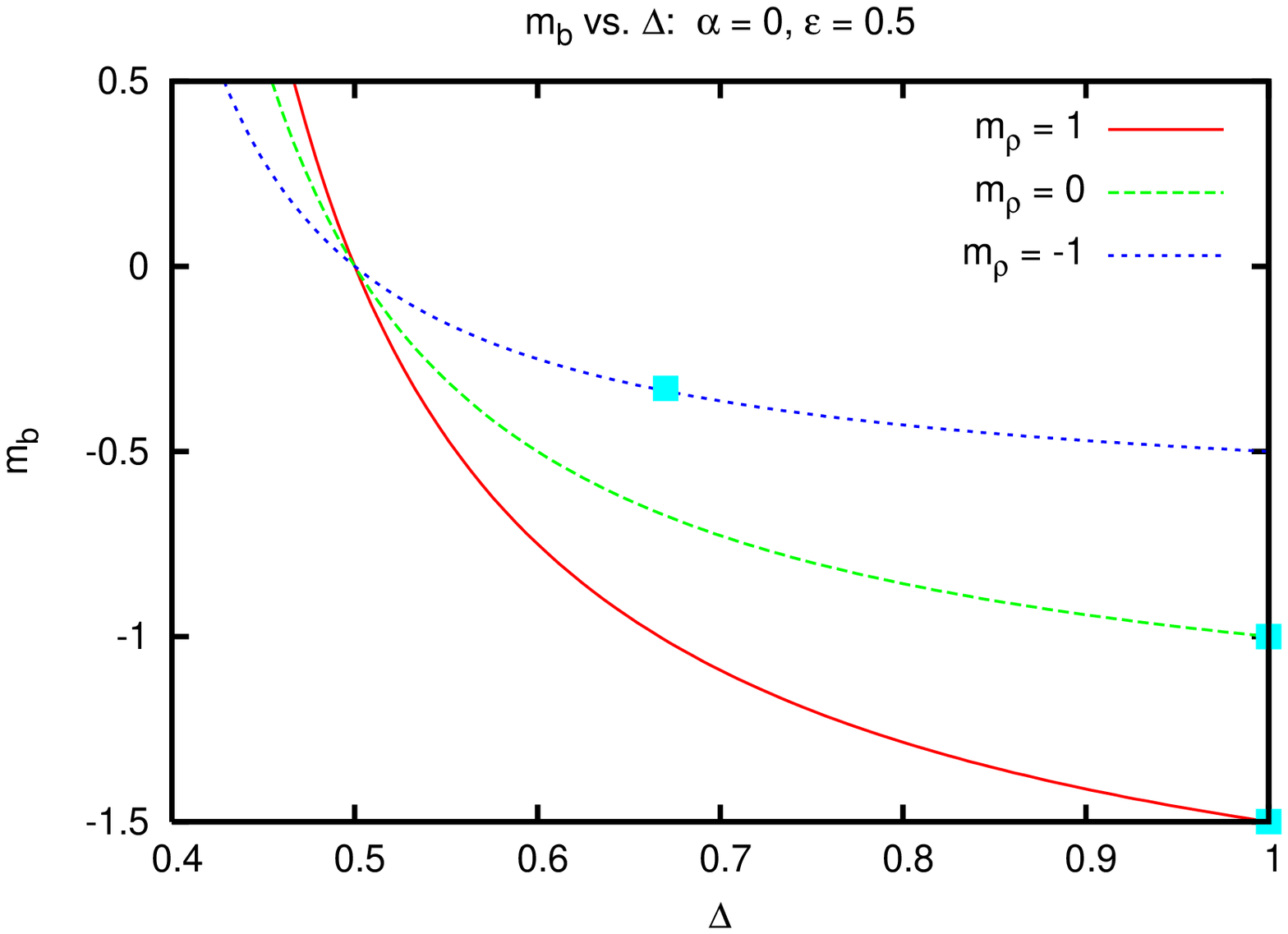}{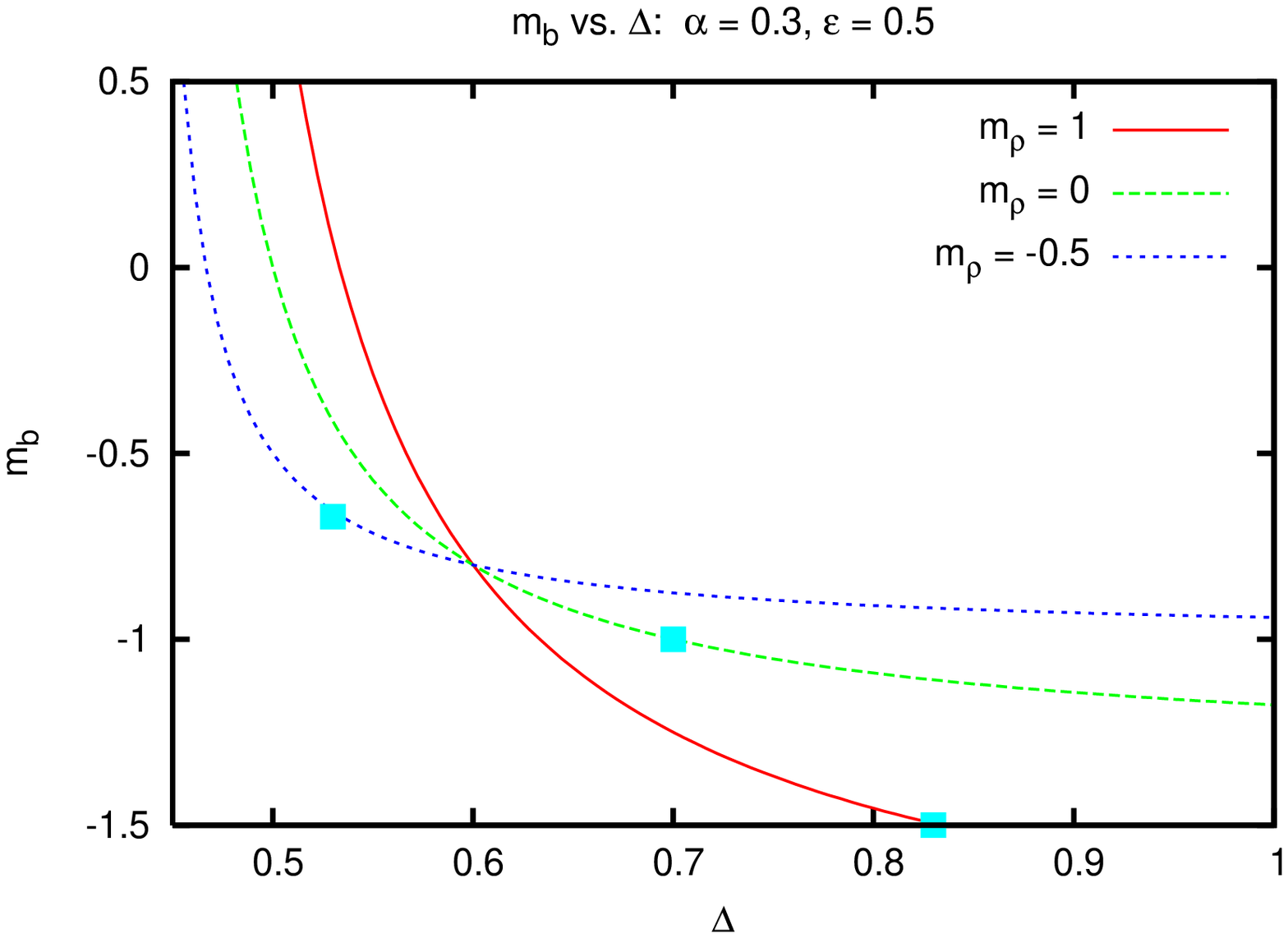}
\caption{As in Figure~\ref{mb1}:  Left, for $\alpha = 0$ and $\epsilon = 0.5$;
right, for $\alpha = 0.3$ and $\epsilon = 0.5$.
\label{mb0p5}}
\end{figure}

Figures~\ref{mb1} and~\ref{mb0p5} plot $m_b$ vs.~$\Delta$ for
Equation~\ref{mb01} and its generalizations to the pairs $(\alpha,
\epsilon) = (0.3, 1),$ $(0, 0.5)$, and $(0.3, 0.5)$, respectively.
Mass conservation is still invoked.  The consistency condition on
$m_c$ places upper limits on $Delta$ (lower limits on $m_b$) shown as
the squares on curves on each plot.  (The condition $1 + 2\epsilon +
m_j > 0$ is met for all curves shown.)  It is difficult to obtain
values of $\Delta > 0.7$; flows with rapidly dropping density (such as
$m_\rho = -2\epsilon$, for constant-velocity mass-conserving flows)
seem unable to do so.  Substantial deceleration seems to be required,
as well as rapid decreases in the magnetic-field strength.  While
there is some parameter space available for accomplishing this,
especially for sources with very flat radio spectra, the most
physically reasonable way to bring about the required deceleration
seems to be mass-loading, which also considerably expands the
available parameter space of source gradients.




\section{Inferring physical parameters}

Energy-loss spectral breaks are commonly used to infer source
magnetic-field strengths in pulsar-wind nebulae.  One requires a
source age $t$; for PWNe in SNR shells, one can use modeling of the
shell emission to estimate an age, while in some cases, a source size
divided by a mean flow speed (estimated one way or another) can give
an age estimate.  Then one simply assumes a homogeneous source for
which $E_c = (a B^2 t)^{-1} = 637/B^2 t$ and from Equation~\ref{ec0},
\begin{equation}
B_h = \left(c_{\rm m} \over a^2\right)^{1/3} \nu_b^{-1/3} t^{-2/3}
= 0.90 \ \nu_{\rm GHz}^{-1/3} \ t_{\rm yr}^{-2/3} \ {\rm G}.
\label{bhom}
\end{equation}
(where we have averaged over pitch angles).  It is of
interest to compare this to the magnetic field that would be inferred
for a flow model assuming (incorrectly) that the source is
homogeneous.  Let the source have a size $L$ and break frequency
$\nu_b$ (so that $L = r_0 l(\nu_b)$).  For a flow model, we can deduce
the initial magnetic field $B_0$ from Equation~\ref{nub} above:
\begin{equation}
\left(L \over r_0\right)^{m_\nu} = {a^2 \over c_{\rm m}} 
 \left(r_0^2 \over v_0^2\right) {1\over f^2} B_0^3 \nu_b
\end{equation}
where $f \equiv 1 + 2m_b -m_v + m_\rho/3$.  This implies
\begin{equation}
B_0 = \left( L \over r_0 \right)^{m_\nu/3} \left(c_{\rm m} \over a^2 \right)^{1/3}
\nu_b^{-1/3} \left(v_0^2 \over r_0^2\right)^{1/3} f^{2/3}.
\end{equation}
Then
\begin{equation}
{B_h \over B_0} = \left(r_0 \over v_0 t\right)^{2/3} f^{-2/3} 
  \left( L \over r_0\right)^{-m_\nu/3}.
\end{equation}

Of course, with substantial magnetic-field gradients, $B_h/B_0$ can
range widely either below or above 1.  Some source properties, such as
the initial ratio of energy input in magnetic field to that in particles
(KC's $\sigma$ parameter), require knowledge of $B_0$.  For those
properties, estimation of source magnetic field from the homogeneous
assumption can lead to significant error.  However, it is possible to
show that $B_h$ does give a good approximation to the mean magnetic
field averaged over the lifetime of a particle moving with the flow, as
of course it must.  The total magnetic-field energy in a flow 
model is
\begin{equation}
U_B = {1 \over 8\pi}\int_1^{L/r_0} r_0 dl \ \pi \left({w(l)\over 2}\right)^2 
    B_0^2 l^{2m_b}
\end{equation}
\begin{equation}
= {1 \over 32} \ \left[{w_0^2 B_0^2 r_0 \over 1 + 2\epsilon + 2m_b} \right] \  
   \left(L \over r_0\right)^{1 + 2\epsilon + 2m_b}
\end{equation}
where we have asumed $1 + 2\epsilon + 2m_b > 0$ and $L \gg r_0$.  The
total volume in the flow is
\begin{equation}
V = \int_1^{L/r_0} \pi \left( {w(l)\over 2}\right)^2 r_0 dl 
   = {\pi r_0 w_0^2 \over 4(1 + 2\epsilon)}
   \left(L \over r_0\right)^{1 + 2\epsilon}.
\end{equation}
Then the mean magnetic energy density $\langle u_B \rangle$ is
\begin{equation}
\langle u_B \rangle \equiv {U_B \over V} = {B_0^2 \over 8 \pi}
   {(1 + 2\epsilon) \over 1 + 2\epsilon + 2m_b} \left(L \over r_0 \right)^{2m_b}
\end{equation}
and the homogeneous energy density $u_B({\rm hom}) \equiv B_h^2/8\pi$ 
satisfies
\begin{equation}
{u_B({\rm hom}) \over \langle u_B \rangle} = 
  f^{-4/3} \left(r_0 \over v_0 t\right)^{4/3} 
  \left(1 + 2\epsilon + 2m_b \over 1 + 2\epsilon\right)
  \left(L \over r_0\right)^{4(1 - m_v)/3}
\label{bratio}
\end{equation}
where the exponent of $L/r_0$ has been rewritten using $m_\nu = 2m_v - 2 - 3m_b$.

If the value of $t$ used in the homogeneous relation
Equation~\ref{bhom} is the actual transit time of an electron from
$r_0$ to $L$, one obtains a similar result.  That $t$ is given by
\begin{equation}
t_{trans} \equiv \int {r_0 dl \over v} = {r_0 \over v_0} \int_1^{L/r_0} l^{-m_v} dl
= {r_0 \over v_0} {1 \over 1 - m_v} \left(L\over r_0\right)^{1 - m_v}.
\end{equation}
Then
\begin{equation}
\left(r_0 \over v_0 t_{\rm trans}\right)^{4/3} \left(L \over r_0\right)^{4(1 - m_v)/3}
  = (1 - m_v)^{4/3}
\end{equation}
independent of physical parameters.  (We are assuming $m_v \le 0$, i.e.,
we exclude {\sl accelerating} flows.) This means that all factors in
Equation~\ref{bratio} are of order unity, so that there is not a large
discrepancy between the true mean magnetic-field energy density and
that inferred under the assumption that the source is homogeneous.
However, if the source lifetime is used for $t$ (which may differ
substantially from $t_{\rm trans}$), or an estimate of $t_{\rm trans}$
is made from the measured expansion velocity of the outer boundary of
the PWN, serious errors may be made in inferring $B$.

\section{Numerical calculations and applications to observed sources}

These results can easily be confirmed by numerical integration of the
appropriate equations.  In particular, Equation~\ref{evoldist} can be
used to find the detailed particle distribution, accounting for the
pileup of particles at energies just below $E_{\rm max}(t)$ as they
migrate down in energy (for $s < 2$).  This pileup can produce a
detectable ``bump'' in the spectrum just below $\nu_b$.  The numerical
calculations can also show how sharp a break can be achieved in
practice.

I illustrate these effects with several models.  The example
parameters for B0540-693 mentioned above $(\alpha = 0.25, \epsilon =
1, m_\rho = 1, m_b = -1, m_v = -2)$ produce distribution functions
$N(E, l)$ at various points in the flow shown in Figure~\ref{b0540},
at positions $l \equiv r/r_0 = 10, 30, 50, 70,$ and $90$.  (The
energies are in units of a fiducial energy $E_f \equiv aB_0^2r_0/v_0$,
the energy an initially infinitely energetic electron would have after
radiating for a time $r_0/v_0$ in a magnetic field $B_0$.)  The sharp
cutoff energy $E_c$, decreasing down the flow, is apparent, as is the
spike just below it of electrons formerly above $E_c.$  The rising
density produces adiabatic {\sl gains} in the density of electrons of
too low energy to be subject to radiative losses. Spatial integration
over these electron distributions produces the model spectral-energy
distribution also shown in Figure~\ref{b0540}, reproduced from
\cite{williams08}, which fits the data surprisingly well, apart from
the anomalous X-ray flux.  (A technical problem, ``pileup'' in the
{\it Chandra} detectors due to the bright X-ray pulsar in B0540-693,
makes the absolute determination of the X-ray flux of the nebula
difficult; see Petre et al.~2007.) For the relatively steep
low-frequency spectrum of B0540-693 ($\alpha = 0.25$), the ``bump''
from integrating over the spikes of Figure~\ref{b0540} is barely
noticeable, but it is much more obvious for a flatter input spectrum.
The model assumes a source radius $L \sim 1.3$ pc \citep{williams08},
but $r_0$ is a free parameter.  If $r_0$ is the pulsar wind
termination shock, we might expect $v_0 = c/3$ \citep{kennel84a}.  The
break frequency (Equation~\ref{nub}) constrains the remaining
combination $r_0^2 B_0^3.$ The model of Figure~\ref{b0540} takes $r_0
= L/100$ and $B_0 = 2.4 \times 10^{-3}$ G.  The radio flux
(Equation~\ref{flux3}) then sets the combination $w_0^2 K_0$; the
model takes $w_0 = r_0/10$ and $K_0 = 2.1 \times 10^{-8}$ cm$^{-3}$
erg$^{0.5}$.  The break frequency calculated from Equation~\ref{nub}
is about $6 \times 10^{12}$ Hz, about a factor of 5 lower than 
the intersection of the extrapolations from low and high frequencies.
This is due to the approximation that electrons radiate entirely
at their peak frequency $\nu_m$, an approximation not made in
the analytic calculations; in general, break frequency predictions
will be low by a factor of several, depending somewhat on the
value of $s$.

\begin{figure}
\epsscale{1.0}
\plottwo{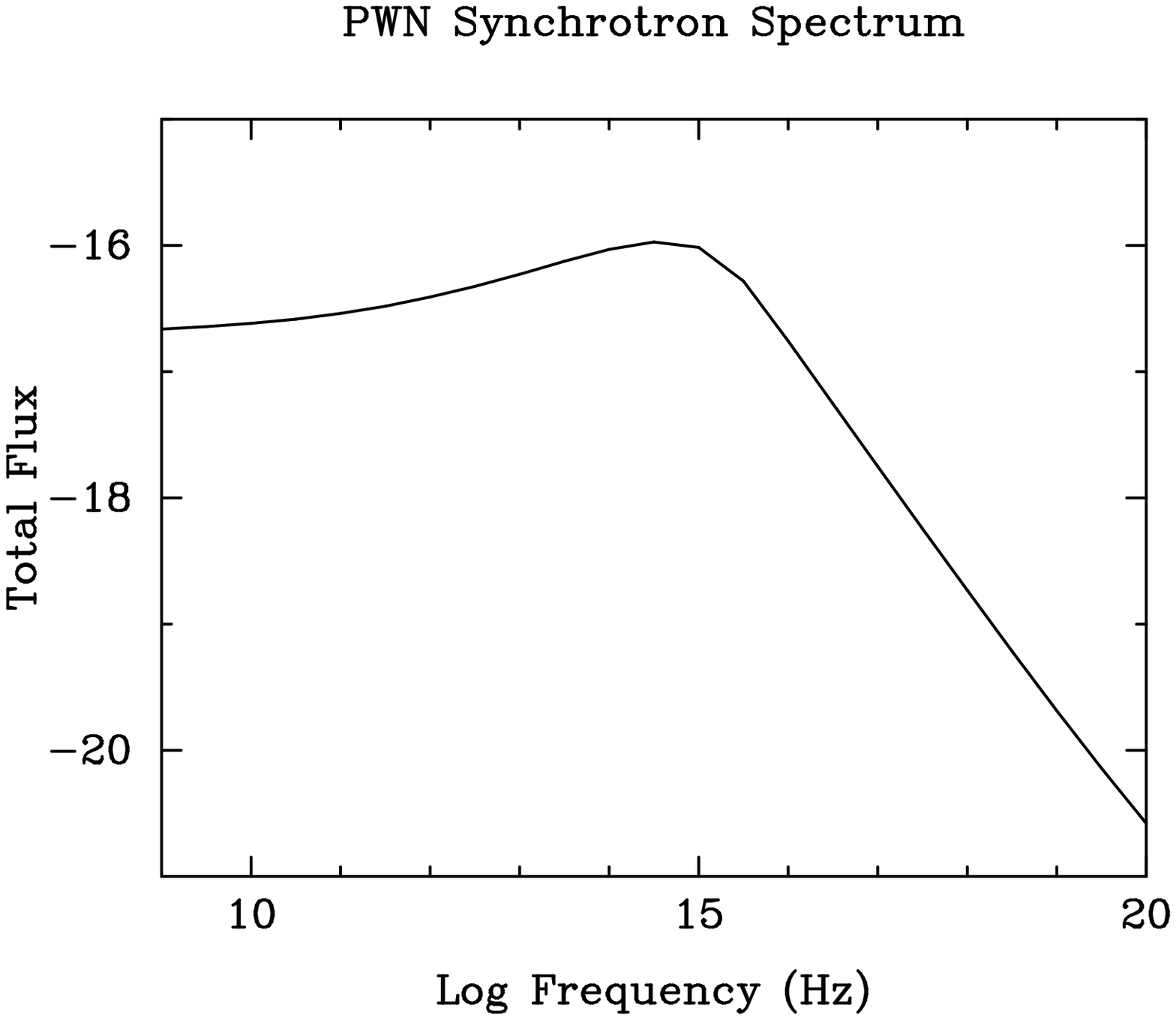}{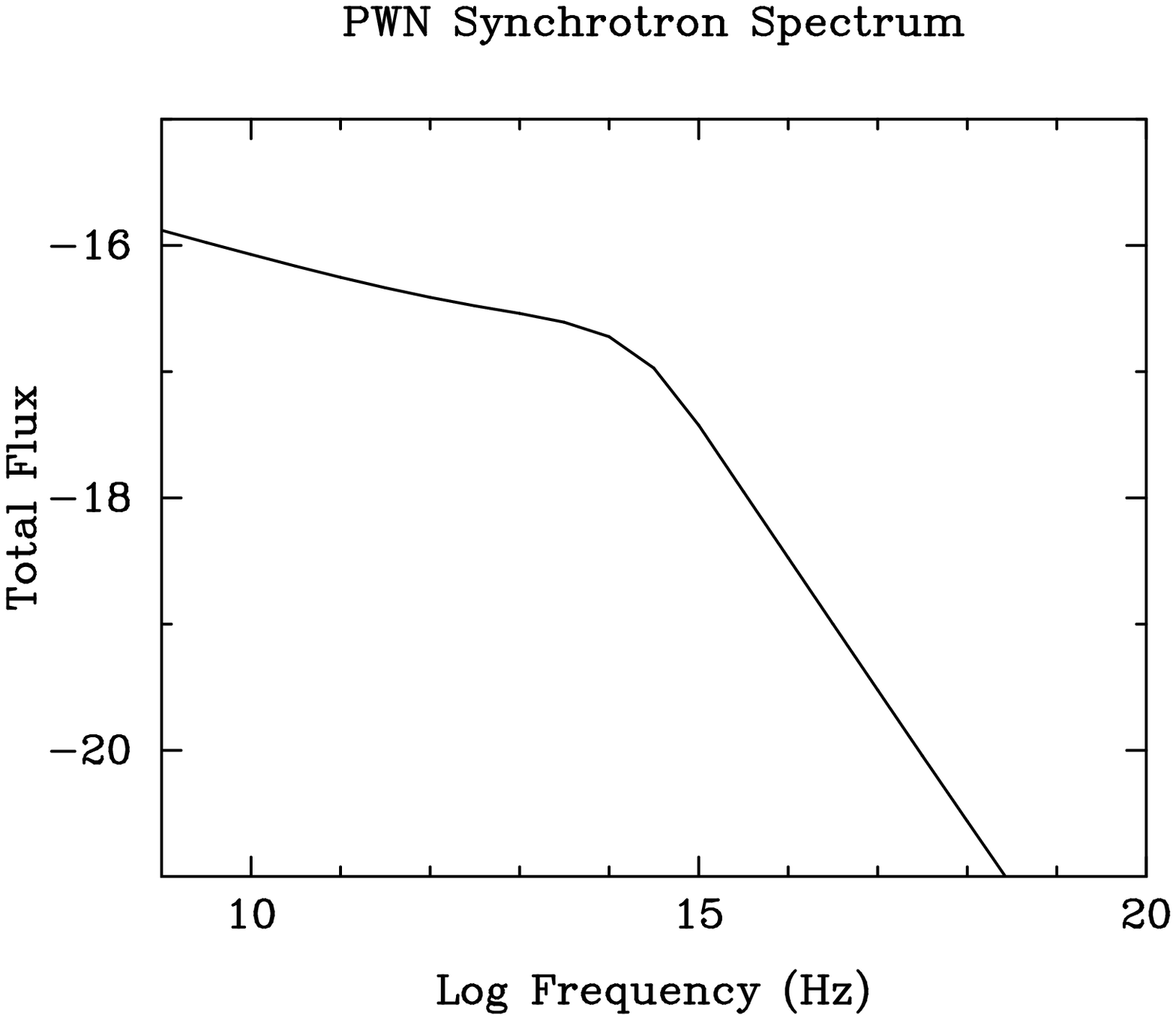}
\caption{Left:  Integrated spectrum for Kes 75 model.  Right:  Same for
MSH 15--5{\it 2}.  Flux scales are arbitrary.  
\label{fluxes}}
\end{figure}

Chevalier (2005) summarizes spectral indices for several PWNe,
including Kes 75 ($\alpha = 0 \Rightarrow s = 0$, $\Delta = 1$) and
MSH 15--5{\it 2} ($\alpha = 0.2 \Rightarrow s = 1.4$, $\Delta =
0.85$).  Such large values of $\Delta$ typically require relaxing
either mass or flux conservation, or both.  For Kes 75, the values
$\epsilon = 1$, $m_\rho = 1$, $m_v = -2$, and $m_b = -1$ (the same as
for the B0540-693 model, except $\alpha = 0$) predict $\Delta = 1.0.$
The consistency condition $1 + 2m_b - m_v > {\rm max}(0, -m_\rho/3)$
is met.  Figure~\ref{fluxes} (left) illustrates the integrated
spectrum.  The ``bump'' is quite prominent; the flux at the peak
around $10^{13.5}$ Hz is 4.4 times that at 1 GHz.  The slope above the
break is $0.96$ between 0.4 and 4 keV, close to the analytic value of
1.0.  Equation~\ref{size} gives the frequency-dependence of the source
size as $l_{\rm max} \propto \nu^{-1/3}$, sufficiently slow that it
might be hard to detect.  For MSH 15--5{\it 2}, Figure~\ref{fluxes}
(right) shows a calculation for $s = 1.4$, $\epsilon = 1$, $m_\rho =
1$, $m_v = -2.22$, and $m_b = -1,$ predicting $\Delta = 0.85.$ The
consistency condition is again met.  The ``bump'' is still
perceptible.  The predicted value of $\Delta$ is reproduced exactly.
The size effect is even slighter: $l_{\rm max} \propto \nu^{-0.29}$.
A factor of 11 frequency range would be required to see the source
shrink by a factor of 2.  For an actual well-resolved source, the
``size'' would need to be measured with some relatively coarse
quantity such as the 50\% enclosed power radius or FWHM, as cited for
the Crab by \cite{kennel84b}.

\section{Summary of results}

Here I collect the principal results and consistency requirements.
The basic result is the expression for $\alpha_2 - \alpha_1 \equiv \Delta$,
the amount of spectral steepening, Equation~\ref{Delta}:
\begin{equation}
\Delta = - {1 + 2\epsilon + m_j \over {m_b + 2m_E}} = 
{1 + 2\epsilon + (2\alpha + 3)m_\rho/3 + (1 + \alpha)m_b
\over 2 + 3m_b -2m_v}. 
\end{equation}
This expression holds if several consistency requirements are met:
$m_E < {\rm min}(0, m_\rho/3)$, where $m_E \equiv -1-2m_b+m_v$, so
that the burnoff energy at position $l$ depends on $l$, and $E_c$
drops with $l$; and $1 + 2\epsilon + (2\alpha + 3)m_\rho/3 + (1 +
\alpha)m_b > 0$, so that the integrated flux density $S_\nu$ depends
on the outer limit of integration.  Finally, the effective source size
should shrink with frequency: $m_\nu < 0,$ a condition always met in
the presence of mass conservation, and almost always met for
reasonable parameters otherwise.  If the conditions are met, the
source size (some measure of the region from which the bulk of the
emission originates) decreases with frequency as
\begin{equation}
l(\nu) \propto \nu^{1/m_\nu} \equiv \nu^{-1/(2 + 3m_b - 2m_v)}.
\end{equation}
if seen more or less from the side; if the flow is nearly along the
line of sight, the expression becomes
\begin{equation}
\theta \propto \nu^{\epsilon/m_\nu} = \nu^{-\epsilon/(2 + 3m_b - 2m_v)}.
\end{equation}
(which is the same for spherical or conical flows where $\epsilon = 1$).

\section{Conclusions}

My basic conclusion is just that synchrotron-loss spectral breaks
differing from 0.5 can be produced naturally in inhomogeneous sources.
I have treated the inhomogeneities resulting from flows, which seem
most natural, using simple power-law parameterizations, but more
complex functional dependencies can be treated the same way.  Other
types of inhomogeneities may be possible as well.  These results are
most straightforwardly applied to PWNe or knots in extragalactic jets,
but may have applications wherever bulk flows of relativistic material
are involved.  In particular, energy-loss breaks seen in gamma-ray
burst afterglows (e.g., Sari et al.~1998; Galama et al.~1998; and much
later work) may provide opportunities for the application of these
results.  For nearby sources, the simplest test of the models is the
detection of the size effect; every model predicts both a particular
$\Delta$ and some rate of decrease of size with frequency (really of
volume, since the size decrease may take place along the line of
sight).  For the same $\Delta$, some variation in the strength of the
size effect is possible.

Fortunately, assuming that an inhomogeneous source is actually
homogeneous does not drastically alter the inferred mean magnetic
field (averaged over the history of a fluid element), but since there
are by assumption large gradients of most quantities, local values of
the magnetic-field strength, such as those at the injection radius,
may depart substantially from the mean values.  This may affect
inferences of the KC magnetization parameter $\sigma$.  Furthermore,
the inference requires knowledge of the actual flow time across the
source -- knowledge that may be hard to come by in the presence of
large velocity gradients.

\acknowledgments

I gratefully acknowledge the hospitality of the Arcetri Observatory of
the University of Florence, where this work was begun.  This work was
also supported by NASA through Spitzer Guest Observer grants RSA
1264893 and RSA 1276758.

\clearpage



\end{document}